# Embracing AI in 5G-Advanced Towards 6G: A Joint 3GPP and O-RAN Perspective

Xingqin Lin, Lopamudra Kundu, Chris Dick, and Soma Velayutham

NVIDIA

{xingqinl, lkundu, cdick, svelayutham}@nvidia.com

*Abstract*—Artificial intelligence (AI) has emerged as a powerful technology that improves system performance and enables new features in 5G and beyond. Standardization, defining functionality and interfaces, is essential for driving the industry alignment required to deliver the mass adoption of AI in 5G-Advanced and 6G. However, fragmented efforts in different standards bodies, such as the third generation partnership project (3GPP) and the open radio access network (O-RAN) Alliance, can lead to confusion and uncertainty about which standards to follow and which aspects of the standards to embrace. This article provides a joint 3GPP and O-RAN perspective on the state of the art in AI adoption in mobile communication systems, including the fundamentals of 5G architecture and its evolution towards openness and intelligence, AI for 5G-Advanced evolution, and a case study on AI-enabled traffic steering. We also identify several areas for future exploration to accelerate AI adoption on the path towards 6G.

## I. INTRODUCTION

We have entered the era of artificial intelligence (AI), brought about by three converging forces, including the availability of big data, the invention of deep learning algorithms, and high-performance computing accelerated by graphics processing units (GPUs) [1]. Meanwhile, the fifth-generation (5G) mobile communication systems are becoming increasingly complex due to the use of advanced technology features and the need to support various services with stringent performance requirements [2]. AI has emerged as a powerful technology that improves system performance and enables new functions in 5G and beyond [3].

The current application of AI in 5G is mainly based on the proprietary implementation in different domains, such as radio access network (RAN), core network, operations support system (OSS), business support system (BSS), and cloud infrastructure [4]. Taking the AI adoption in 5G and beyond to the next level requires industry alignment through global standardization efforts. Indeed, a variety of AI-related activities have been taking place in many standardization bodies, including the third generation partnership project (3GPP) and the open RAN (O-RAN) Alliance, among others [5].

While standardization is crucial for industry alignment to increase AI adoption in 5G and beyond, fragmented efforts in different standards bodies can lead to confusion and uncertainty about which standards to follow and which aspects of the standards to embrace. Therefore, it is essential to develop a holistic view of the diverse AI-related activities in different standards bodies and connect them organically. However, existing works either scratch the surface of the AI-related activities across different standards bodies without sufficient depth [5] or only focus on the AI-related activities in a particular standards body [6].

This article aims to address the gap in the existing literature by providing a joint 3GPP and O-RAN perspective on AI adoption in 5G networks, focusing on the RAN aspects. Given that the standards work on AI adoption is carried out within the established 5G architecture, we first describe the split RAN architecture developed by 3GPP to lay the foundation for the subsequent discussion. We then introduce the further disaggregated architecture specified by O-RAN, which infuses openness and intelligence into 5G architecture. After that, we provide an overview of the 3GPP work on AI adoption in the 5G evolution, including AI-enabled RAN intelligence and AI for the new radio (NR) air interface. We also present a case study on traffic steering, which has attracted interest from both 3GPP and O-RAN, to illustrate how the different aspects of the 3GPP and O-RAN standards work on AI adoption fit together. We conclude this article with a 6G outlook by pointing out fruitful research avenues to accelerate AI adoption on the path to 6G.

## II. FUNDAMENTALS OF 5G ARCHITECTURE

The architecture of 5G radio access network (RAN) (aka. next-generation RAN (NG-RAN)) is the basis for incorporating AI in 5G networks. In this section, we describe the preliminaries of NG-RAN architecture [7] to pave the way for the subsequent discussions on how AI can be incorporated within the NG-RAN framework.

The basic building block of the NG-RAN architecture is the NG-RAN node, which is a logical node defined by a set of logical functions and the associated logical interfaces that it terminates towards other logical nodes. An NG-RAN node can either be a 5G node B (gNB) providing new radio (NR) access, or a next-generation enhanced node B (ng-eNB) providing long-term evolution (LTE) access. In this article, we focus on NG-RAN consisting of a set of gNBs. The gNBs are connected to 5G core (5GC) network through the NG interface. They can also be interconnected through the Xn interface. Figure 1 provides an illustration of the overall 3GPP-defined NG-RAN architecture [7]. The logical architecture accommodates diverse deployment options (e.g., centralized, distributed, and monolithic) in a transparent manner.

To support different deployment migration paths, NG-RAN supports two modes of operation: non-standalone (NSA) operation and standalone (SA) operation. In NSA operation, gNBs and ng-eNBs are connected to the same core network and interoperate to provide dual connectivity to UE. Dual connectivity allows the UE to use resources provided by two different nodes: master node and secondary node. A set of architecture options can be defined by combining different combinations of gNB or ng-eNB as master node or secondary node and different core network types (LTE core network and 5GC).

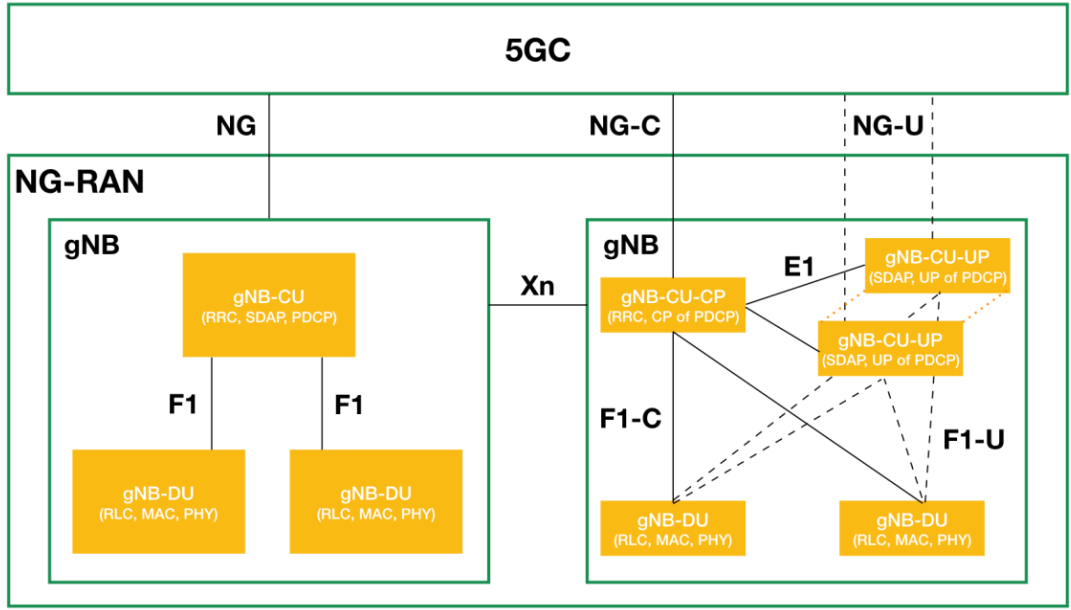

**Figure 1 An illustration of NG-RAN architecture**

A gNB can be split into a central unit (gNB-CU) and one or more distributed units (gNB-DU), which are logical nodes. A gNB-CU and a gNB-DU are connected through the F1 interface. The gNB-CU hosts radio resource control (RRC) protocol, service data adaptation protocol (SDAP), and packet data convergence protocol (PDCP) of the gNB. The gNB-DU hosts radio link control (RLC), medium access control (MAC), and physical (PHY) layers of the gNB. One gNB-DU can support one or multiple cells. The gNB-CU manages UE context and instructs the gNB-DU on the radio resource allocation for the UE. The gNB-DU is responsible for the scheduling of the radio resources.

A gNB-CU can be further split into a control plane part (gNB-CU-CP) and one or more user plane parts (gNB-CU-UP), which are logical nodes. This allows for further optimization of distributing different RAN functions. A gNB-CU-CP and a gNB-CU-UP are connected through the E1 interface. The gNB-CU-CP hosts the RRC and the control plane part of the PDCP of the gNB-CU. It terminates the control plane part of F1 interface (F1-C) towards a gNB-DU. The gNB-CU-UP hosts the SDAP and the user plane part of the PDCP of the gNB-CU. It terminates the user plane part of F1 interface (F1-U) towards a gNB-DU. The gNB-CU-CP can request the gNB-CU-UP to set up, modify, and release the bearer context, which includes information about data radio bearers, protocol data unit (PDU) sessions and quality of service (QoS) flows associated with a UE. The gNB-CU-UP is responsible for informing the gNB-CU-CP of UE's inactivity and reporting data volume to the gNB-CU-CP.

In summary, NG-RAN features NSA and SA operation modes, a multitude of architecture options with dual connectivity, and split gNB architecture. The split gNB architecture brings more flexibility in implementation, facilitates traffic load management, enables virtualized deployments, and allows better scalability. The flexibility of the NG-RAN architecture makes it future-proof and forms the basis for adopting AI in 5G RAN, which will be discussed in more detail in the subsequent sections.

### III. EVOLUTION OF 5G ARCHITECTURE TOWARDS OPENNESS AND INTELLIGENCE

With the vision of disaggregating NG-RAN towards an open, virtualized, interoperable, and AI driven architecture, O-RAN Alliance augmented the CU/DU-split framework with a set of open interfaces [8]. These interfaces include both the enhanced 3GPP interfaces (F1, E1, Xn, NG) as well as additional interfaces standardized by O-RAN (A1, E2, open fronthaul (FH), O1, O2) interconnecting logical/physical O-RAN nodes. Figure 2 shows the high-level O-RAN-defined architecture with key nodes and interfaces enabling disaggregation of NG-RAN.

Within the O-RAN infrastructure, RAN domain management services including fault, configuration, accounting, performance, and security (FCAPS) are provided by service management and orchestration (SMO) framework. It contains a non-real-time (non-RT) RAN intelligent controller (RIC) function and interfaces with other O-RAN network functions (O-NFs) through A1, O1, open FH management (M)-plane, and O2 interfaces. Key O-NFs orchestrated by SMO include near-real-time (near-RT) RIC, O-RAN central unit- control plane/user plane (O-CU-CP/UP), distributed unit (O-DU), and radio unit (O-RU). These O-RAN components can either be realized through virtualized network functions (VNFs), e.g., virtual machines (VMs) or containers hosted by O-RAN cloud

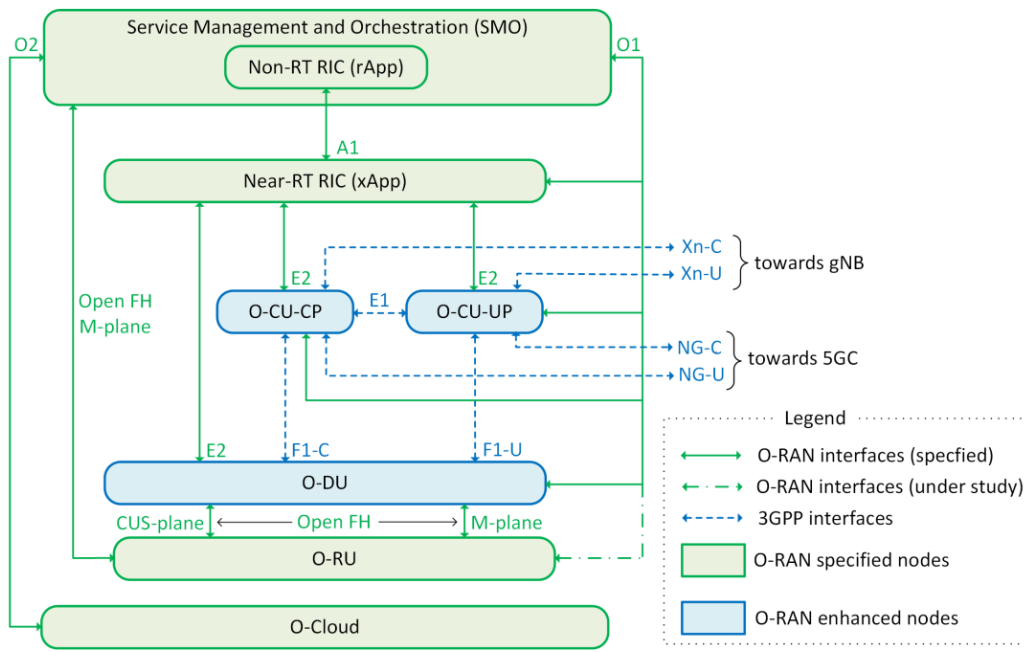

**Figure 2 An illustration of O-RAN logical architecture**

(O-Cloud) platform; or through physical network functions (PNFs) utilizing specialized hardware.

Native to SMO, non-RT RIC function terminates A1 interface towards near-RT RIC and aids intelligent RAN optimization by providing policy-based guidance, machine learning (ML) model management and enrichment information to near-RT RIC. In addition, non-RT RIC applications (rApps) perform data-driven radio resource optimization in non-real time (exceeding one second) by leveraging SMO services and functionalities exposed by non-RT RIC framework, and triggering actions through A1, O1, O2 and open FH M-plane interfaces.

Near-RT RIC is a logical function in O-RAN architecture, enabling near-real-time (~10 ms to 1 s) control and optimization of O-RAN NFs terminated by E2 interface. In the purview of NG-RAN architecture, near-RT RIC serves as an overlay, rendering AI-driven capabilities and action-triggers to NG-RAN nodes (such as CUs and DUs) over O-RAN specified E2 interface. Similar to the application-based service model of non-RT RIC (rApps), near-RT RIC hosts a set of xApps that anchor to E2 interface for near-real-time data collection (UE specific or cell-specific) and in turn, provides value added services to E2 nodes (logical nodes terminating E2 interface). Further details of RIC functionalities and associated service models are discussed in [9].

Southbound of near-RT RIC are E2 nodes, viz., O-CU and O-DU. Following the control/user plane flow separation model of NG-RAN architecture, O-CU is further partitioned into two logical nodes: O-CU-CP and O-CU-UP. These nodes represent the counterparts of gNB-CU-CP and gNB-CU-UP respectively in the O-RAN architecture. In particular, gNB-CU-CP/UP interfaces (F1, E1, Xn , NG) and associated protocols extend to O-CU-CP/UP nodes but are adapted to O-RAN specified interoperability profiles to enable multi-vendor solution. Additionally, O-CU-CP/UP terminate E2 and O1 interfaces towards near-RT RIC and SMO, respectively.

The functionality of the gNB-DU in NG-RAN manifests into two separate nodes in O-RAN architecture: O-DU (logical node) and O-RU (physical node). The decoupling follows 7-2x lower layer split (LLS), one of the various functional split options studied by 3GPP in the past. As a sub-variant of intra-PHY split (Option 7), 7-2x aggregates RLC, MAC, and the majority of the PHY functionalities (High-PHY) within O-DU, while allocating the rest of the PHY functionalities (Low-PHY), radio frequency (RF) processing, and transmission-reception points (TRPs) to O-RU. O-RAN Alliance adopted 7-2x split considering trade-offs between several factors (e.g., O-RU cost/complexity, fronthaul bandwidth requirement, achievable latency and throughput, control signaling overhead) to achieve multi-vendor interoperable RAN [10].

Within the O-RAN framework, disaggregated O-DU and O-RU interconnect over open FH interface, consisting of control, user, synchronization (CUS) plane and M-plane [10]. Support of O-RU management by the O-DU is enabled by hierarchical or hybrid mode. In hybrid mode, SMO also connects with O-RU over M-plane for FCAPS functionality, while in hierarchical mode, O-DU exclusively manages the O-RU. Both O-DU and O-RU terminate O1 interface towards SMO.

O-NFs (such as near-RT RIC, O-CU-CP/UP, and O-DU) are hosted by O-Cloud platform as cloud-native network functions (CNFs) or physical network functions (PNFs). O-Cloud comprises of a set of physical infrastructure nodes supporting necessary software components (e.g., operating systems, virtual machine monitor, container runtime), commodity hardware (e.g., hardware accelerators such as GPUs, field programmable gate arrays (FPGAs), and application-specific integrated circuit (ASICs), network switches, network interface cards (NICs)) and a collection of management/orchestration functions to export O2 interface for RAN workload management on cloud.

In addition, O-Cloud supports a set of internal acceleration abstraction layer (AAL) application programming interfaces (APIs) to disaggregate RAN software from underlying hardware platform - a key feature towards enabling CU/DU application portability across multi-vendor RAN components. Virtualization of NG-RAN with O-Cloud makes network deployment options more versatile and facilitates remote lifecycle management (LCM) including software upgrades, zero touch configuration and automation.

To recap, O-RAN architecture infuses openness and intelligence into 5G and beyond, evolving NG-RAN towards a multi-vendor infrastructure based on the principles of flexibility, scalability, virtualization, softwarization, energy efficiency, and most importantly, 'RAN intelligence'- embedding AI/ML capability within the RAN.

## IV. AI FOR 5G-ADVANCED EVOLUTION

In parallel to the O-RAN Alliance's work on evolving NG-RAN architecture to embed AI capabilities, 3GPP has also been investigating the incorporation of AI in the 5G evolution, including AI for NG-RAN and AI for NR air interface.

### A. AI for NG-RAN in 3GPP Release 17

3GPP conducted a study on AI-enabled RAN intelligence in Release 17 [11]. A set of high-level principles were identified to guide the standards work. One key principle is that the detailed AI algorithms and models for use cases are implementation specific and thus are not expected to be standardized. This principle aims to avoid over-specifying AI in 5G standards to foster competition and innovation. Therefore, the study focused on AI functionality and the corresponding types of inputs and outputs instead of the detailed AI algorithms and models. For example, the standards work can focus on data collection to provide the required information to the model training or inference function (e.g., on a subscription basis) and leaves the details of how the model training or inference function uses the data to implementation. Respecting user data privacy is essential in data collection.

The study on AI-enabled RAN intelligence defined a reference functional framework. Common terminologies such as data collection, model training, model inference, and actor are introduced as part of the functional framework. This facilitates industry alignment around the basic architecture and concepts of AI-enabled RAN intelligence, which is indispensable to realizing a multivendor ecosystem. However, it should be emphasized that the functional framework serves as a reference and does not mandate all use cases to follow it strictly.

One guiding principle of the standards work is that the many aspects of AI-enabled RAN intelligence, such as the input and output of the model training or inference function and where the model training or inference function resides, should be studied on a case-by-case basis. For each use case, different deployment options are possible for the AI functions. In one example, the AI model training function can reside in the operations, administration, and maintenance (OAM), and the AI model inference function can reside in the gNB (or gNB-CU for split gNB). Alternatively, both AI model training and inference functions can reside in the gNB (or gNB-CU for split gNB). The study on AI-enabled RAN intelligence focused on three use cases: network energy saving, load balancing, and mobility optimization, which are anticipated to bring substantial benefits in evolved RAN.

*Network energy saving:* The rapid growth of 5G deployment has led to concerns of high energy consumption, carbon emissions, and high operation cost. Network energy saving is a complicated problem involving multiple layers of the network and the need to balance with other key performance indicators (KPIs) and QoS requirements. Conventional network energy saving mechanisms rely on rule-based configuration, e.g., switching on/off cells based on different thresholds of cell load. Such rule-based techniques are reactive, inflexible, and difficult for achieving globally optimized system performance and energy efficiency. AI algorithms can leverage the RAN data to optimize network energy saving, e.g., predicting energy efficiency and load in future states to enable proactive, adaptive actions of traffic offloading, coverage modification, and cell activation/ deactivation.

*Load balancing:* Steering traffic to balance the load in 5G networks becomes increasingly more challenging due to the use of multiple frequency bands and interworking with different radio access networks. Distributing load or traffic offloading in 5G networks is usually achieved by optimizing handover parameters and decisions. The existing rule-based load balancing decisions (primarily relying on the current or historical load information) have difficulty in coping with fast time-varying scenarios with high mobility and dynamic traffic patterns of a large number of connections. AI algorithms can leverage the collection of RAN data such as measurements and feedback from network nodes and UEs to predict load to improve network performance and user experience.

*Mobility optimization:* Mobility support is a distinct feature of mobile communication systems, including 5G. Mobility management ensures mobile users' service continuity by minimizing radio link failures and handover disruptions. Mobility optimization encounters increasing complexity in 5G, which needs to support high-mobility UEs with stringent QoS requirements (e.g., low latency, high reliability, and zero interruption) in high-frequency networks with ever-smaller and more irregular cells. Furthermore, mobility optimization needs to address advanced features in 5G, such as sophisticated dual connectivity options, conditional handover, and dual active protocol stack (DAPS) handover. AI algorithms can leverage the collection of RAN data to improve handover performance, predict UE location and performance, and steer traffic to achieve quality network performance.

For each use case, the study on AI-enabled RAN intelligence identified a set of input information, a set of output information, and a set of feedback information. The input information may be generated by different entities, including the local NG-RAN node hosting the AI inference function, neighboring NG-RAN nodes, and UEs. Examples of the input information are RAN energy efficiency metrics, RAN resource status measurements, UE location information, UE mobility history, UE radio measurements, UE traffic information, and UE performance information. The output information includes a variety of predicted performance metrics and recommended actions. Examples of the output information are recommended cell activation/deactivation strategy, recommended target cells to offload traffic, predicted energy efficiency, predicted resource status information, predicted UE trajectory, and predicted UE traffic. The set of feedback information is used to derive training or inference data or monitor the AI model's performance for each use case. Examples of the feedback information are RAN resource status, system KPIs, and UE performance information. Figure 3 shows a high-level reference signaling flow for AI model training and model inference in a NG-RAN node. It illustrates the flow of the input information, output information, and feedback information.

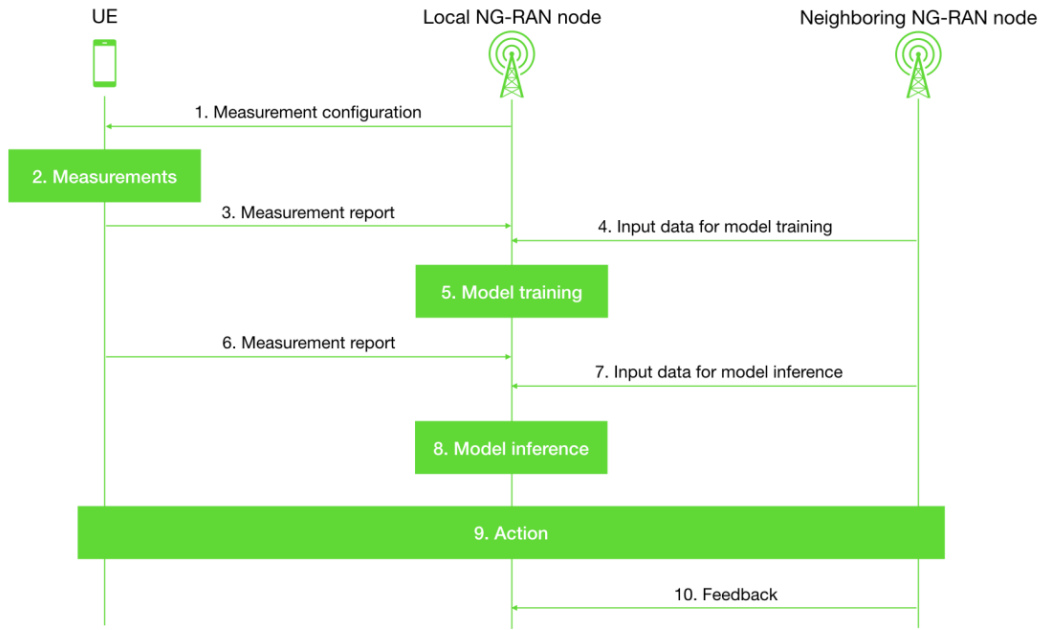

**Figure 3 A high-level reference signaling flow for AI model training and model inference in a NG-RAN node**

*B. AI for NG-RAN and NR Air Interface in 3GPP Release 18*

With the completion of the Release-17 study on AI-enabled RAN intelligence, 3GPP is working on the normative work in Release 18 to specify data collection enhancements and signaling support for AI-based network energy saving, load balancing, and mobility optimization [12]. It is anticipated that the standardization will introduce features to facilitate event prediction and make use of predictions, which is a leap from the measurement-centric processes in the existing 5G networks.

In Release 18, 3GPP is not only conducting normative work on AI-enabled RAN intelligence but also exploring the use of AI to augment the NR air interface, which is the first of its kind in the evolution of cellular networks [12]. The objective of the study is to set up a general framework for enhancing air interface with AI capabilities. A comprehensive list of topics is under exploration, including defining stages of AI algorithms, gNB-UE collaboration levels, needed datasets for AI model training, validation, and testing, life cycle management of AI models, etc. Similar to the study on AI-enabled RAN intelligence, the study on AI for NR air interface adopts a use-case-centric approach, focusing on three selective use cases: channel state information (CSI) feedback enhancement, beam management enhancement, and positioning accuracy enhancement. Unlike the study on AI-enabled RAN intelligence which did not discuss detailed AI algorithms and models, the study on AI for NR air interface defines evaluation mythology and KPIs for conducting performance evaluations of AI algorithms and models, despite that the details of the AI algorithms and models are not expected to be standardized. The study also assesses the potential specification impact from the physical layer to higher layer protocols to interoperability and testability, which will pave the way for future normative work.

V. USE CASE: AI-ENABLED TRAFFIC STEERING

Efforts of evolving various features and components of open RAN architecture are spanned across multiple technical working groups (WGs) and task groups (TGs) under the umbrella of O-RAN Alliance. The specification works ongoing in these disjoint groups are harmonized towards a key goal - enabling a set of potential use cases for open RAN deployment. Towards that goal, O-RAN WG1 (use cases and overall architecture working group) identifies potential use cases, defines high-level use case details in terms of architectural requirements and workflows, and disseminates the analysis to relevant working groups within O-RAN for inter-WG alignment towards use case enablement. Use case down selection and prioritization are influenced by ecosystem contributions and feedback (e.g., through operator surveys), coordination with other standard development organizations (SDOs) relevant for O-RAN work and research on emerging industry trends towards next generation open RAN, amongst various factors.

In the context of cross-SDO influence, some of the use cases analyzed by O-RAN are interrelated with the AI-enabled RAN intelligence use cases studied by 3GPP, as described in the previous section. One prominent example in this category is the O-RAN defined use case of 'traffic steering'. The key goal of this use case enablement is to allow network operators to specify various objectives related to traffic management, such as improving network efficiency (e.g., by energy saving), achieving proportionate cell load (i.e., load balancing), or optimizing UE performance (e.g., mobility optimization). Thanks to the native AI/ML capability of O-RAN architecture, operators can dynamically configure desired network policies, set the targeted performance criteria, and leverage ML capabilities to achieve agile, intelligent, and intent-based traffic management.

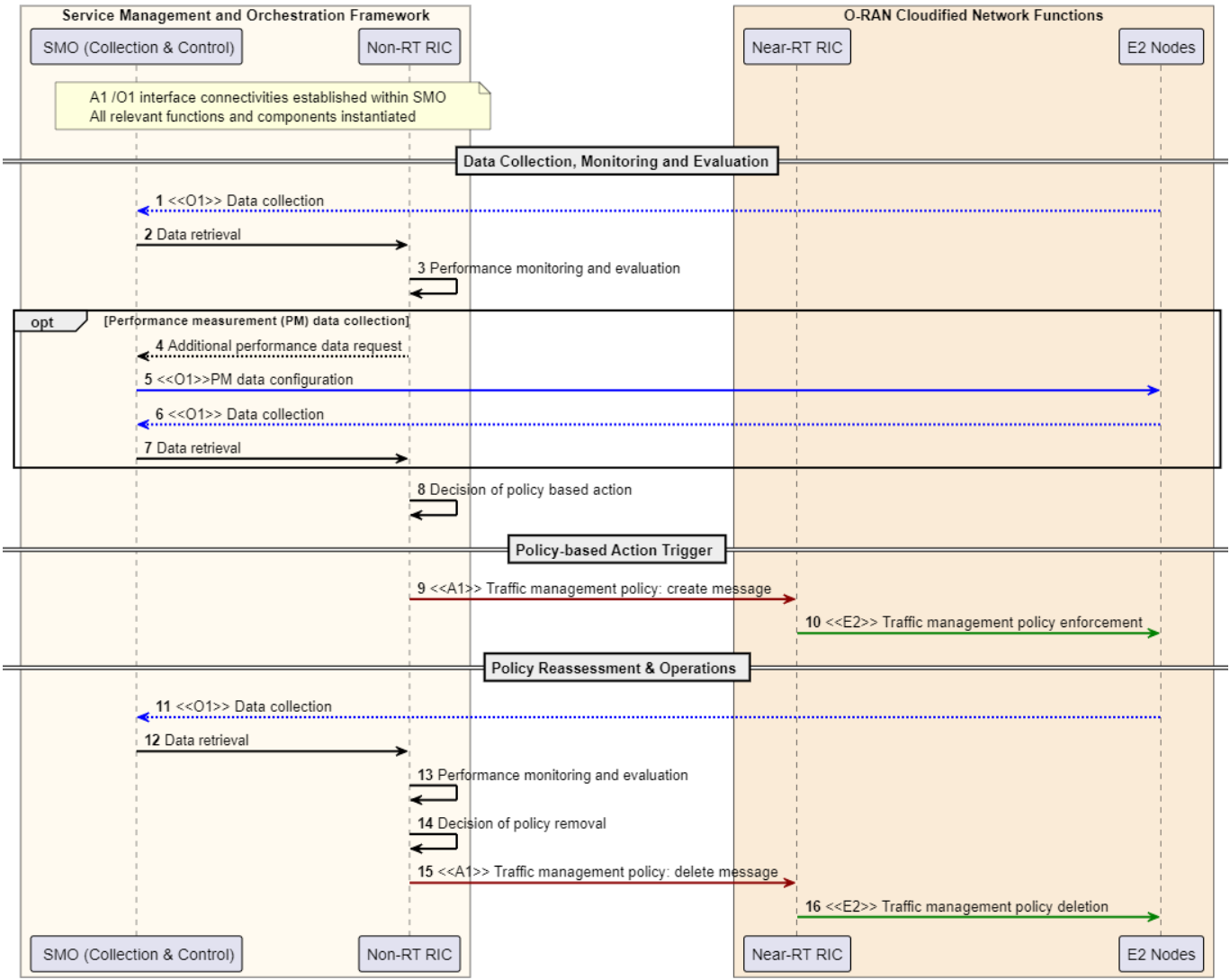

**Figure 4 Flow diagram for RIC enabled traffic steering in O-RAN architecture**

Figure 4 illustrates the flow diagram envisioned for enabling traffic steering use case in O-RAN architecture [13]. The primary entities involved in this workflow are SMO (collection & control), non-RT RIC, near-RT RIC, E2 nodes (O-CU/O-DU), and associated interfaces (O1, A1, E2). While SMO serves as the termination point for O1 interface, non-RT RIC and near-RT RIC act as RAN policy control and RAN policy enforcement functions respectively, with E2 nodes providing control and user plane functionalities. Prior to triggering traffic steering operations, SMO establishes data collection and sharing process, ensuring that non-RT RIC has access to the collected data. In particular, non-RT RIC monitors network performance through relevant measurements and counters collected from E2 nodes via SMO over O1 interface. In an operational network, use case specific call flows are typically triggered upon the detection of a set of operator-specified conditions or events. At the onset of the trigger for traffic steering, a series of sequential steps occur to complete three key phases of operations, viz., 1) data collection, monitoring and evaluation, 2) policy-based action trigger, and 3) policy reassessment and operations.

In the first phase, non-RT RIC retrieves necessary data from SMO, including (optionally) additional measurement data and counters specifically collected from E2 nodes by SMO as per non-RT RIC's configuration request, and assesses network performance. Based on the outcome of performance monitoring and evaluation, non-RT RIC determines a policy-triggering action.

In the second phase, non-RT RIC communicates relevant policies to near-RT RIC over A1 interface, which may include QoS target, cell allocation preference for control/user planes, and aspects of bearer handling such as bearer selection and bearer type change. On receiving the information from non-RT RIC, relevant xApps running on near-RT RIC interpret the specified policy and derive tangible actions to be enforced on E2 nodes.

In the third phase, non-RT RIC continues monitoring and evaluating network performance post policy implementation by collecting relevant performance events and counters from E2 nodes via SMO. During this phase, non-RT RIC may decide at some point of time that the conditions to continue the policy is no longer valid. If that happens, non-RT RIC would trigger a following action for policy removal. On receiving relevant message, near-RT RIC subsequently would delete the policy previously instantiated on E2 nodes.

AI-enabled traffic steering use case aggregates measurement data and KPIs by different grouping types, e.g., based on cells, QoS types, and RAN slices. The required data is primarily based on 3GPP-defined metrics such as measurement reports for serving and neighboring cells (reference signal received power (RSRP), reference signal received quality (RSRQ), channel quality indicator (CQI), etc.), intra and inter-radio access technology (RAT) measurement reports (for multi-access scenarios), cell global identifier (CGI) reports, measurement gaps (per-UE or per-frequency), and the like. Additional performance statistics tailored towards this specific use case are augmented by O-RAN, including UE connection and mobility handover statistics (with handover status indication - success/failure), cell load statistics such as number of active users and scheduled users per transmission time interval and frequency resource utilization, per user performance statistics like PDCP throughput and RLC/MAC layer latency.

## VI. CONCLUSION AND 6G OUTLOOK

The fast growth of the 5G network scale and the diverse application demands require intelligent solutions to manage the ever-increasing complexity. AI has emerged as a powerful technology that can enable network automation, boost system performance, and improve user experience. Standardization, defining functionality and interfaces, is essential for driving the industry alignment required to deliver the mass adoption of AI in mobile communication systems. This article has provided a joint 3GPP and O-RAN perspective on AI adoption in 5G networks, focusing on the RAN aspects.

Though the journey of AI adoption in mobile communication systems has begun in 5G, it has come only as an afterthought within the already established 5G architecture and air interface. 6G research is under way, and 6G standardization is expected to start in 3GPP around 2025. We will have an opportunity to design the 6G system by embracing AI natively right from the beginning. The journey towards large-scale AI adoption in 6G should be taken with a joint effort across the standards bodies to avoid diverging directions and industry fragmentation.

We conclude by pointing out some fruitful research avenues to accelerate AI adoption on the path to 6G.

*Digital twins*: A digital twin network is a virtual representation of a physical network. It features closed-loop interaction between virtual and real networks. Digital twins can help accelerate AI adoption towards 6G. One key area is data generation. AI-native 6G inherently hinges on the availability of data sets for training, validation, and testing. Although mobile communication systems generate vast amounts of data, their general availability is limited. A digital twin network built in a platform such as Omniverse [14] can generate high-fidelity synthetic data, covering rare but crucial corner cases, for developing AI algorithms. Another key area is the assessment of AI algorithms in the digital twin network before deploying them in the real network. This is a critical capability because assessing AI algorithms in the real network can be risky and may jeopardize the normal operation of the real network.

*Converged communication and computing architecture:* Native AI in 6G will span across all network layers, requiring first class support of AI training and inference processing across cloud, core, RAN, and device. It is vital that the AI systems can learn continuously (e.g., online learning) to adapt to varying environments, site-specific conditions, and heterogenous configurations. The underlying platforms must be flexible to support diverse AI applications as well as wireless signal processing and communication [15]. To this end, a softwarized approach based on programmable computing hardware, such as GPU, can play a crucial role in accelerating AI adoption on the path to 6G. A key research area will be to explore and advance the technologies needed to support the convergence of communication and computing architecture.

*Trustworthy AI*: It is anticipated that 6G will provide services for virtually every part of life, society, and industry. For such critical 6G systems, trustworthiness is an essential requirement. To realize the full potential of AI in 6G, it is crucial to ensure that humans can understand AI applications and functionalities, such as how the AI algorithms make certain predictions, decisions, and recommendations, and consequently trust their use at all levels of 6G. As an example, explainable AI can provide the needed interpretability and transparency of the AI algorithms. Research advancements with trustworthy AI for 6G will facilitate its adoption at scale.